\documentclass[reprint,superscriptaddress, amsmath,amssymb,pra, aps,floatfix,longbibliography]{revtex4-2}
\usepackage[english]{babel}

\usepackage{hyperref}
\usepackage{bm}
\usepackage{hyperref}
\usepackage{graphicx}
\usepackage[capitalise]{cleveref}
\usepackage{textcomp}
\usepackage{xfrac}
\usepackage{flushend}

\hypersetup{pdftoolbar=true,        
    pdfmenubar=true,        
    pdffitwindow=false,     
    pdfstartview={FitH},    
    pdftitle={Kerr Microresonator Soliton Frequency Combs at Cryogenic Temperatures},    
    pdfauthor={Gregory Moille},     
    pdfsubject={CryoComb},   
    pdfcreator={Gregory Moille},   
    pdfproducer={LaTeX}, 
    pdfkeywords={Version 1.0},
    colorlinks=true,       
    linkcolor=blue,          
    citecolor=blue,        
    filecolor=blue,      
    urlcolor=blue}
\newcommand{\beginsupplement}{%
        \setcounter{table}{0}
        \renewcommand{\thetable}{S\arabic{table}}%
        \setcounter{figure}{0}
        \renewcommand{\thefigure}{S\arabic{figure}}%
        \renewcommand{\thesubsection}{S-\arabic{subsection}}
     }

\newcommand{\ks}[1]{#1}
\begin{document}
\title{Dissipative Kerr Solitons in a III-V Microresonator}

\author{Gregory Moille}
  \thanks{These authors contributed equally}
  \affiliation{Microsystems and Nanotechnology Division, National Institute of Standards and Technology, Gaithersburg, USA}
  \affiliation{Joint Quantum Institute, NIST/University of Maryland, College Park, USA}
\author{Lin Chang}
  \thanks{These authors contributed equally}
  \affiliation{Department of Electrical and Computer Engineering, University of California, Santa Barbara, USA}

\author{Weiqiang Xie}
  \thanks{These authors contributed equally}
  \affiliation{Department of Electrical and Computer Engineering, University of California, Santa Barbara, USA}
\author{Ashutosh Rao}
  \affiliation{Microsystems and Nanotechnology Division, National Institute of Standards and Technology, Gaithersburg, USA}
  \affiliation{Institute for Research in Electronics and Applied Physics and Maryland Nanocenter, University of Maryland,College Park, USA}
\author{Xiyuan Lu}
  \affiliation{Microsystems and Nanotechnology Division, National Institute of Standards and Technology, Gaithersburg, USA}
  \affiliation{Institute for Research in Electronics and Applied Physics and Maryland Nanocenter, University of Maryland,College Park, USA}
\author{Marcelo Davanco}
  \affiliation{Microsystems and Nanotechnology Division, National Institute of Standards and Technology, Gaithersburg, USA}
\author{John E. Bowers}
  \email{bowers@ece.ucsb.edu}
  \affiliation{Department of Electrical and Computer Engineering, University of California, Santa Barbara, USA}
\author{Kartik Srinivasan}
  \email{kartik.srinivasan@nist.gov}
  \affiliation{Microsystems and Nanotechnology Division, National Institute of Standards and Technology, Gaithersburg, USA}
  \affiliation{Joint Quantum Institute, NIST/University of Maryland, College Park, USA}

\date{\today}

\begin{abstract}
    We demonstrate stable microresonator Kerr soliton frequency combs in a III-V platform (AlGaAs on SiO$_2$) through quenching of thermorefractive effects by cryogenic cooling to temperatures between 4~K and 20~K. This cooling reduces the resonator's  thermorefractive coefficient, whose room-temperature value is an order of magnitude larger than that of other microcomb platforms like Si$_3$N$_4$, SiO$_2$, and AlN, by more than two orders of magnitude, and makes soliton states adiabatically accessible. Realizing such phase-stable soliton operation is critical for applications that fully exploit the ultra-high effective nonlinearity and high optical quality factors exhibited by this platform.
\end{abstract}

\maketitle

Micoresonator frequency combs based on dissipative Kerr solitons (DKSs)~\cite{kippenberg_dissipative_2018} are promising for chip-scale metrology implementation, including applications such as optical clocks~\cite{Ludlow2015,Drake:2018da4}, spectroscopy systems~\cite{picque2019frequency}, and range measurements~\cite{Suh884}. DKSs have been demonstrated in amorphous dielectrics such as SiO\textsubscript{2}~\cite{Yi2015} and Si\textsubscript{3}N\textsubscript{4}~\cite{Gaeta_soliton_Si3N4}, crystalline materials such as MgF\textsubscript{2}~\cite{Kippenberg_soliton_2014}, III-Nitride materials such as AlN~\cite{liu_beyond_2019,Gong:18}, and promising new integrated photonics platforms such as thin film LiNbO\textsubscript{3}~\cite{He:19}. In contrast, III-As materials, though they are particularly attractive for microcomb applications due to the ability to simultaneously realize ultra-high effective nonlinearity~\cite{pu2016efficient} and large optical quality factor~\cite{chang2019ultra,wilson2018gallium}, are yet to show stable microcavity DKS operation. Given the competing processes that can occur in such materials, including thermorefractive, free carrier, and photorefractive effects, achieving this milestone is important to further establish the potential of these platforms for applications that require phase-stable frequency comb generation. III-V materials, mostly binary or ternary compounds made of Aluminum (Al), Gallium (Ga), Arsenic (As) and Phosphorus (P), started to gain attention more than three decade ago for photonics applications thanks to their large third-order nonlinearity~\cite{Ho:91, Aitchison:97} - two-to-three orders of magnitude higher than SiO\textsubscript{2}~\cite{ikeda2008thermal}, Si\textsubscript{3}N\textsubscript{4}~\cite{ikeda2008thermal} and AlN~\cite{Jung:13} - and the avoidance of two-photon absorption (TPA) in the telecom band~\cite{STEGEMAN:94} thanks to band-gap engineering~\cite{CAPASSO172}. Reduction in TPA resulted in several milestones for nonlinear photonics such as waveguide soliton pulse compression in suspended GaAs photonic crystal membranes~\cite{colman2010temporal} , energy-efficient optical gates~\cite{moille2016integrated}, and ultra-low threshold optical parameter oscillators~\cite{martin2016triply}. A major breakthrough has been the emergence of the III-V on insulator platform\cite{pu2016efficient}, which has enabled strong modal confinement and dispersion control, and for which longstanding challenges with respect to scattering losses have been mitigated~\cite{Deri:87},%
resulting in the realization of high quality facto($Q>10^6$) microring resonators~\cite{chang2019ultra,wilson2018gallium} in which geometric dispersion control can be exercised. Together with the ultra-high effective nonlinearity ($\gamma = 377$~W\textsuperscript{-1}m\textsuperscript{-1}), these devices have achieved comb generation at pump powers easily reachable with chip-integrated lasers~\cite{chang2019ultra}. However, while a transient 'soliton step' was observed in Ref.~\cite{chang2019ultra}, indicating soliton existence, stable access and operation on a DKS state was not achieved in that work.
\begin{figure*}
    \centering
    \includegraphics[width=\textwidth]{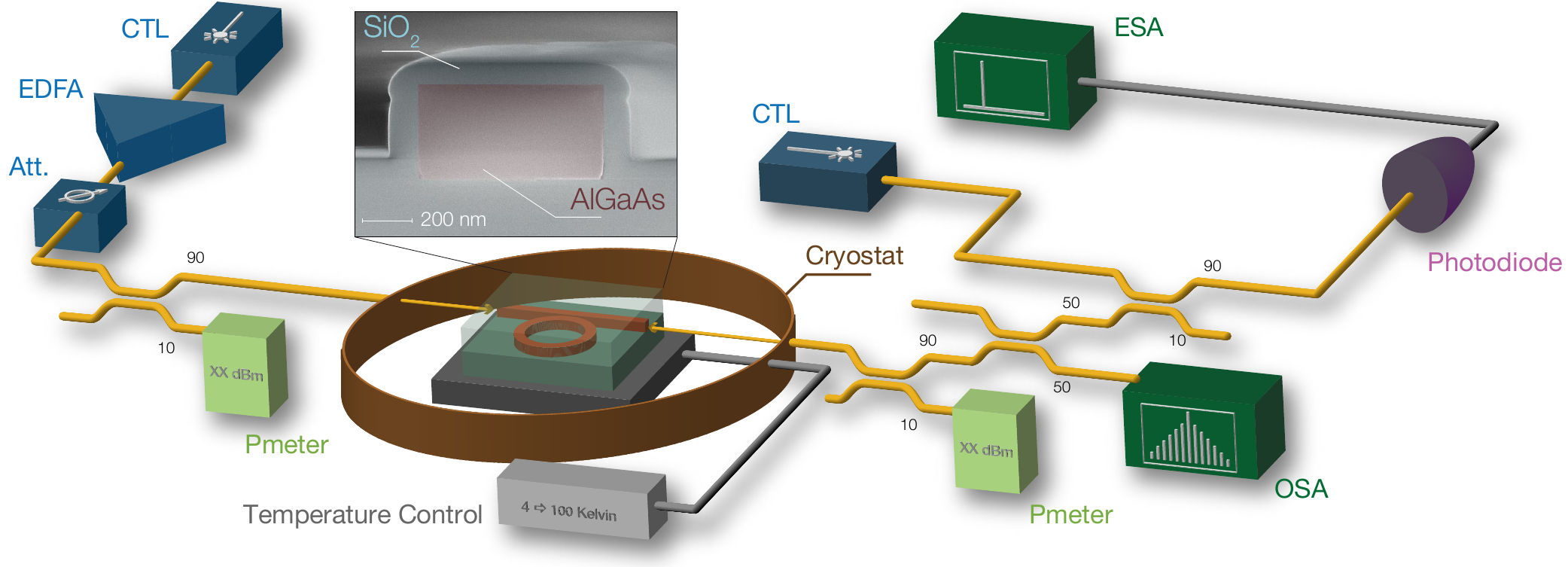}
    \caption{\label{fig:Setup}Experimental setup for generating a DKS in an AlGaAs-on-insulator microring resonator, where the sample is placed on top of a temperature-controlled mount inside a closed-cycle cryostat to modify the resonator's thermorefractive coefficient and achieve conditions favorable for DKS generation. The soliton is generated by manually tuning the frequency of the CTL pump laser, while the beat note measurement is performed using a second CTL laser. CTL: Continuous Tunable Laser, EDFA: Erbium Doped Fiber Amplifier, Pmeter: Power Meter, OSA: Optical Spectra Analyzer, ESA: Electrical Spectra Analyzer}
\end{figure*}

\indent ‘Here, we demonstrate the first stable generation of DKSs in an ultra-high nonlinearity III-V microresonator. The DKS is thermally accessible under manual tuning of the pump laser frequency, due to a dramatic drop in the thermorefractive coefficient ($\partial{n}/\partial{T}$, where $n$ is the system's refractive index and $T$ is temperature) realized by a cryogenic operating temperature ($T<20$~K). We measure the system's temperature-dependent $\partial{n}/\partial{T}$ and show that its large gradient creates a limited window for which DKSs are thermally accessible. Within this range there is a balance between high enough power to sustain the DKS and low enough power to avoid strong heating that increases the resonator's thermorefractive coefficient and reduces the DKS accessibility region (i.e., shortens the DKS step length).\\
\indent The experimental setup (\cref{fig:Setup}) is similar to ref.~\cite{PhysRevApplied.12.034057}, where the sample is in a 20~cm diameter closed-cycle cryostat whose sample mount temperature can be set from 4~K to 100~K. The resonator is a microring of radius $R=18.97$~$\mu$m and ring width $RW=740$~nm, made in 400~nm thick Al\textsubscript{0.2}Ga\textsubscript{0.8}As encapsulated in SiO\textsubscript{2}~\cite{chang2019ultra}, \ks{and the resulting calculated dispersion characteristics are given in the Supplementary Materials.} The microring chip is accessed by lensed optical fibers (insertion loss of 6 dB/facet), and is pumped by an amplified continuous tunable laser (CTL). The resonator output spectrum is monitored with an optical spectrum analyzer (OSA), and a second CTL is used in heterodyne beat note measurements against the generated comb, detected by a 12~GHz photodiode and an electrical spectral analyzer.\\

\begin{figure}[b]
    \centering
    \includegraphics[width = \columnwidth]{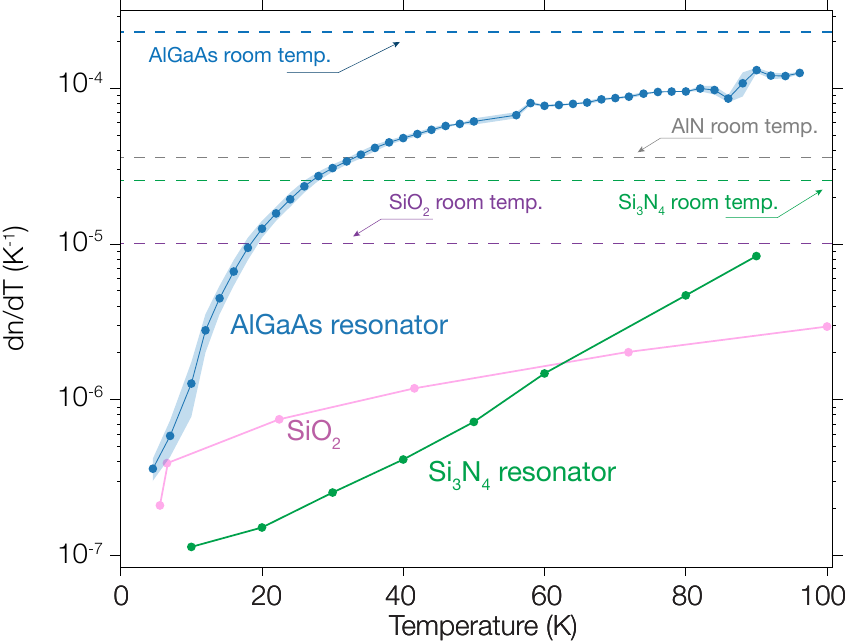}
    \caption{\label{fig:dndT}Measured thermorefractive coefficient $\partial{n}/\partial{T}$ versus $T$ for the AlGaAs resonator (blue), Si\textsubscript{3}N\textsubscript{4} resonator from ref.~\cite{PhysRevApplied.12.034057} (green), SiO\textsubscript{2} from ref.~\cite{elshaari_thermo-optic_2016} (pink) and the room-$T$ values for AlGaAs (dashed blue line) from ref.\cite{camargo20042d}, Si$_3$N$_4$ (dashed green line) from ref.~\cite{PhysRevApplied.12.034057}, and AlN (dashed grey line) from ref.~\cite{Watanabe_2012}. The light blue area shows a range of $\partial n/\partial T$ values for the AlGaAs resonator if the sample temperature differs from the sample holder by up to $\pm 1$~K.}
\end{figure}
\indent First, we measure the shift of the resonances with temperature ($T$) to extract the thermorefractive coefficient $\partial{n}/\partial{T}$, where we probe the frequency shift of a resonance - the one which will be further pumped to generate the soliton frequency comb around 191.4~THz - while changing the temperature of the sample mount from 7~K to 96~K. We do not observe significant change in either the coupling $Q$ (average $\overline{Q_\mathrm{c}}=850\times10^3$) or the intrinsic $Q$ (average $\overline{Q_\mathrm{i}}=575\times10^3$), as shown in S2. We believe that the resonator is limited by scattering losses and because the microring is pumped far from its band-edge, we cannot draw conclusions about a potential change of absorption due to temperature-induced band-gap modification. In order to retrieve the temperature dependence of $\partial n/ \partial T$ from the measured frequency shift, we first calculate $\partial \nu / \partial n = -55.98$~THz from an eigenmode solver (see S3), where $\nu$ is the eigenfrequency, and $n$ is the refractive index of the AlGaAs, with the assumption that any variation in the refractive index of SiO\textsubscript{2} has little effect on the resonance frequency due to its small value and the large modal confinement within the AlGaAs layer (see S3). We then retrieve $\partial n / \partial T = \left(\partial \nu/\partial n \right)^{-1} \partial \nu/\partial T$%
, with the results displayed in \cref{fig:dndT}. It is important to note that we disregard thermal expansion as a potential source of a temperature-dependent resonance frequency shift, given its comparatively small expected value (near 10$^{-7}$ K$^{-1}$), as reported in the literature~\cite{Novikova}.\\
\indent The results of the measurements in \cref{fig:dndT} show that in addition to a nearly $100$ times higher thermorefractive coefficient than Si\textsubscript{3}N\textsubscript{4} and SiO\textsubscript{2} between 30~K and 60~K, the slope of $\partial n/\partial T$ (\textit{i.e.} $\partial^2 n/\partial T^2$) for the AlGaAs system is quite large, as evidenced by an order of magnitude change in the range between 7~K$<T<$20~K. This suggests a more limited pump power and temperature window for soliton accessibility under slow pump frequency tuning in AlGaAs than in Si$_3$N$_4$~\cite{PhysRevApplied.12.034057}.\\
%
%
\begin{figure}[t]
    \centering
    \includegraphics{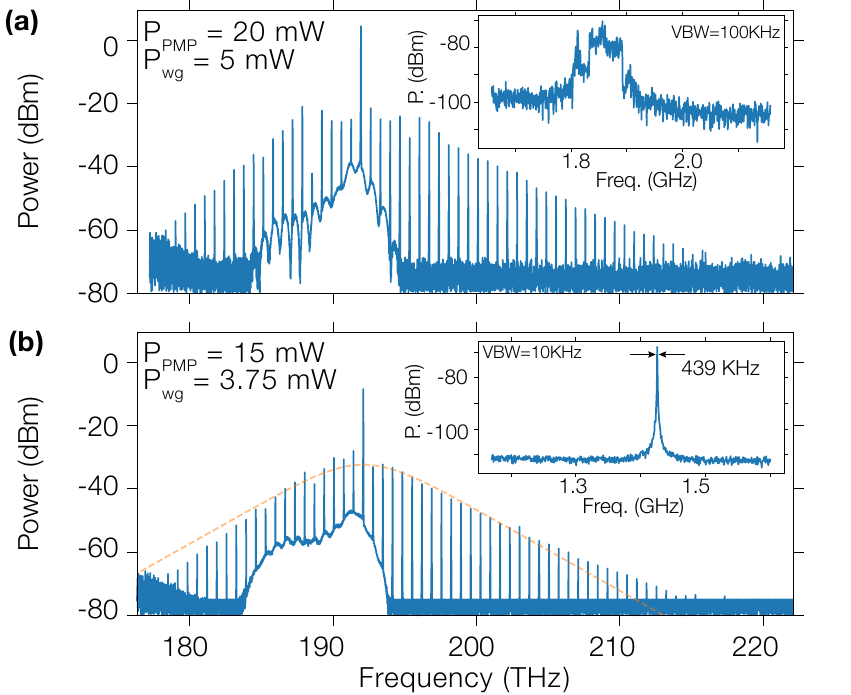}
    \caption{\label{fig:Spectra}(a) MI comb spectrum for the resonator pumped at 5~mW in-waveguide power. (b) Single soliton spectrum when pumped at 3.75~mW in-waveguide power. Insets display beat note spectra acquired through interference of the comb tooth closest to 197~THz with a CW laser. These spectra are taken from DC to 6.8~GHz; the displayed data are zoomed in on the region of interest.}
\end{figure}
\begin{figure}[t]
    \centering
    \includegraphics[width = \columnwidth]{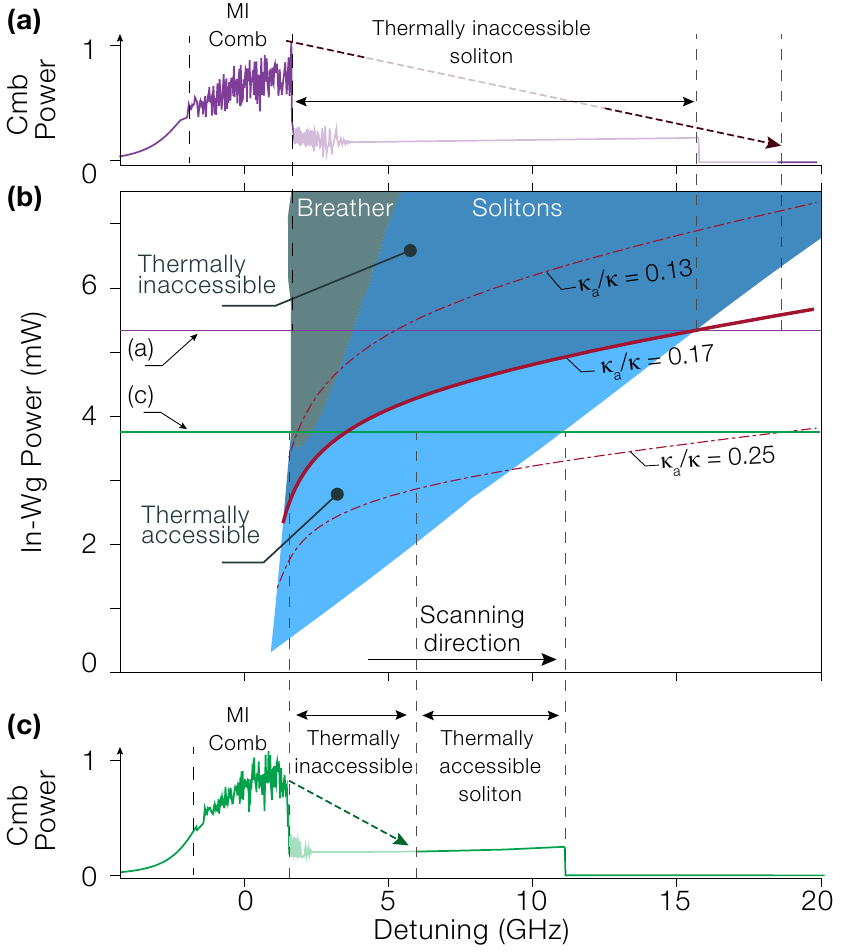}
    \caption{\label{fig:Stability} Soliton accessibility diagram highlighting the impact of the large gradient in $\partial n/\partial T$ between 4~K to 20~K. (a) High pump power heats up the device, resulting in a large $\partial n/\partial T$ and no adiabatically accessible soliton. (c) Lower pump power generating a comb while keeping the temperature low enough for an adiabatically accessible soliton. Panel (b) shows the soliton region (blue) computed from the Lugiato-Lefever equation (\textit{i.e.} only Kerr), while the solution \ks{to the} thermal model is plotted as \ks{red traces for three fractional absorption coefficients. For a given power, the soliton region to the left of these traces is not thermally accessible. The shaded region to the right of the traces corresponds to the thermally accessible solitons in our experiment. The light shaded blue region is accessible for the case of $\kappa_{a}/\kappa$=0.17.}
    These regions are qualitatively defined, highlighting the crux of the experimental observations, but full quantitative detail is not achieved due to uncertainties in parameters that enter into the thermal model. The purple and green lines correspond to panels a and c, respectively, and are placed at their corresponding pump powers.}
\end{figure}
\indent In~\cref{fig:Spectra} we examine comb generation by pumping the resonator at $f_\mathrm{pmp}=$191~THz, increasing the pump power to the few mW level, \ks{and sweeping the pump frequency with the laser piezo element manually, hence orders of magnitude slower than the expected thermal time constant (expected to be on the order of tenths of a $\mu$s)}. When the pump power is too high (in-waveguide power $>$5~mW), 
only modulation instability (M.I.) states are observable, resulting in a comb spectrum that significantly differs from the $sech^2$ envelope expected for a soliton state.  This is consistent with the broad beat note (inset to \cref{fig:Spectra}(a)) obtained through interference of the comb and a stabilized tunable laser near 197~THz. In contrast, for an in-waveguide power between 3~mW and 5~mW \ks{(about 10~$\times$ greater than the estimated threshold power),} the resulting comb spectrum (\cref{fig:Spectra}(b)) is significantly different from the M.I one, and follows the expected $sech^2$ envelope. Furthermore, the beat note measurement (inset to ~\cref{fig:Spectra}(b)) shows a single narrow peak with significantly reduced noise floor relative to the M.I. state. This data, taken together, is indicative of stable access to a DKS state. \ks{It is interesting to note the comb tooth around 187~THz appears to be lower than the comb envelope in both the M.I. and DKS states, which indicates a possible avoided mode crossing. However, it is spectrally separated far enough from the pump that its influence on DKS formation is apparently limited.}
The limited power window over which soliton states are adiabatically accessible appear to be strongly linked to the thermal properties of the system. Indeed, Al\textsubscript{0.2}Ga\textsubscript{0.8}As presents a relatively good thermal conductivity ($k_\mathrm{AlGaAs} = 62.2$~W.m\textsuperscript{-1}.K\textsuperscript{-1}~\cite{Afromowitz:73}), and is about a factor of 2 largen than Si\textsubscript{3}N\textsubscript{4} ($k_\mathrm{Si\textsubscript{3}N\textsubscript{4}} = 30 $~W.m\textsuperscript{-1}.K\textsuperscript{-1}~\cite{ikeda2008thermal}). However, the microring in this study is being embedded within SiO\textsubscript{2} which is a thermal insulator ($k_\mathrm{SiO\textsubscript{2}} \simeq 1$~W.m\textsuperscript{-1}.K\textsuperscript{-1}~\cite{Yamane:2002}).%

As a result, the temperature of the resonator can increase by a few degrees once a sufficiently large circulating power is reached. This relatively small increase of resonator temperature due to the pump will result in a large increase of the thermorefractive coefficient, resulting in a soliton state that is no longer thermally accessible. \cref{fig:Stability}(b) summarizes this physical picture in a qualitative way, \ks{using the model described in the Supplement} (uncertainty in quantities such as the resonator absorption rate and heat capacity preclude a fully quantitative description). In this figure, the soliton existence region within the 2D landscape of in-waveguide pump power and pump laser detuning from the cavity mode has been computed using the steady-state Lugiato-Lefever equation through the \textit{pyLLE} software~\cite{Moille:2019291}, which only includes Kerr-mediated effects. Here, we have defined detuning as positive when the laser is on the red-detuned side of the cavity (i.e., at lower frequency). Thermal accessibility is addressed by a secondary equation as described in ref.~\cite{Li2017}, and is directly proportional to the thermorefractive index. Its solution is depicted by the red solid line in \cref{fig:Stability}(b). If the pump power is low enough, for example, corresponding to the horizontal green line in \cref{fig:Stability}(b), $\partial n /\partial T$ is low enough, and the soliton existence window extends in laser detuning beyond the thermally inaccessible region and can be accessed adiabatically through slow frequency tuning of the pump (\cref{fig:Stability}(c)). However, when the pump power is too high (purple horizontal line in \cref{fig:Stability}(b)), the soliton existence window lies entirely within a thermally inaccessible region and can not be accessed adiabatically (\cref{fig:Stability}(a)), resulting solely in the observation of MI states.


%
%

\indent In conclusion, we have demonstrated stable DKS operation in a III-V microring resonator. This has been realized by reducing its thermorefractive coefficient by more than two orders of magnitude through operation at cryogenic temperature. We further show that due to the large gradient of the thermorefractive coefficient between 4~K and 20~K, the pump power window over which soliton states are thermally accessible is limited. Our results show that thermorefractive effects have been the main limitation to soliton accessibility at room temperature, and point to the importance of minimizing optical absorption \ks{and maximizing thermal conductivity. }%
\ks{Such reduction in absorption could be obtained by better surface passivation~\cite{moille2016integrated,Guha:17,chang2019ultra}, while thermal mitigation can be provided using other cladding material~\cite{Zheng:19}. Together with effective pump-cavity detuning methods like use of an auxiliary laser~\cite{Zhang:19} or integrated heaters~\cite{stern2018battery}, this may open the possibility of realizing straightforward room-temperature access to DKS in a III-V platform.}\\

\noindent \textbf{Acknowledgments.} The authors acknowledge support from the DARPA DODOS program. G.M and K.S. thank Alfredo de Rossi for fruitful discussions.
\vspace{-0.2in}

\bibliographystyle{lpr}
\bibliography{Biblio}
clearpage
\section*{Supplementary Material}
\beginsupplement

\subsection{Dispersion}

The dispersion is computed using an eigeinfrequency solver taking into account the chromatic dispersion of the AlGaAs and the SiO2. The real part of the dielectric constant of AlGaAs, hence its refractive index, can be retrieved using the formulas developed by S. Adachi~\cite{Sadao} such that:

\begin{align*}
    \varepsilon_\mathrm{Al_{x}Ga_{1-x}As}(\nu) =& A_0(x)\left( f(\xi) + \frac{1}{2}\left( \sfrac{E_0}{E_0 + \Delta_0} \right)^{\sfrac{3}{2}})f(\xi_0)\right) \\
                                              & + B_0(x)
\end{align*}

\noindent where $\nu$ is the optical frequency, $x$ ($1-x$) the Al fraction (Ga fraction) in the ternary compound, $\xi = \sfrac{h\nu}{E_0}$, $\xi_0 = \sfrac{h\nu}{E_0 + \Delta_0}$ , $A_0(x) = 6.3 + 19x$, $B_0(x) = 9.4 + 10.2x$, $E_0 = 1.425 + 1.155x +0.37x^2$~(units of eV), $E_0+\Delta_0 = 1.765 + 1.11x + 0.37x^2$~(units of eV), $h$ is the Planck constant, and the function $f(\xi)$ is :

\begin{equation*}
    f(\xi) = \xi^{-2}\left(2 - (1+\xi)^{\sfrac{1}{2}} - (1-\xi)^{\sfrac{1}{2}}\right)
\end{equation*}

\noindent The dielectric constant of SiO$_2$ is given by a Sellmeier equation as~\cite{Malitson:65}:
\begin{equation*}
    \varepsilon_\mathrm{SiO_2}(\lambda_\mu) = 1+\sum_{i=1}^{3} \frac{A_i \lambda_\mu^2}{\lambda_\mu^2-B_i}
\end{equation*}

\noindent with $A_1 = 0.6961663$, $B_1 =0.0684043$, $A_2 = 0.4079426$, $B_1 =0.1162414$, $A_3 = 0.8974794$, $B_3 =9.896161$ and $\lambda_\mu$ the wavelength in microns.

\begin{figure}[h]
    \centering
    \includegraphics[width=\columnwidth]{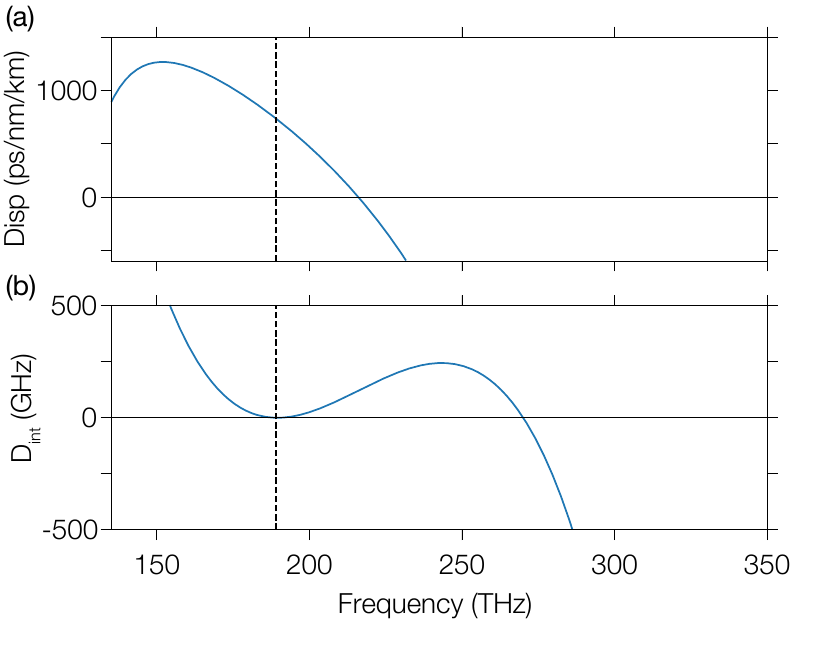}
    \caption{\label{fig:DispDint} (a) Dispersion parameter and (b) Integrated dispersion of an AlGaAs ring resonator embedded in SiO$_2$ and with the following geometrical parameters: $RW=740$~nm, $H=400$~nm, and $R=18.97$~µm. The dashed line corresponds to the 191~THz pump frequency used in the experiment.}
\end{figure}

\begin{figure}[h]
    \centering
    \includegraphics[width=\columnwidth]{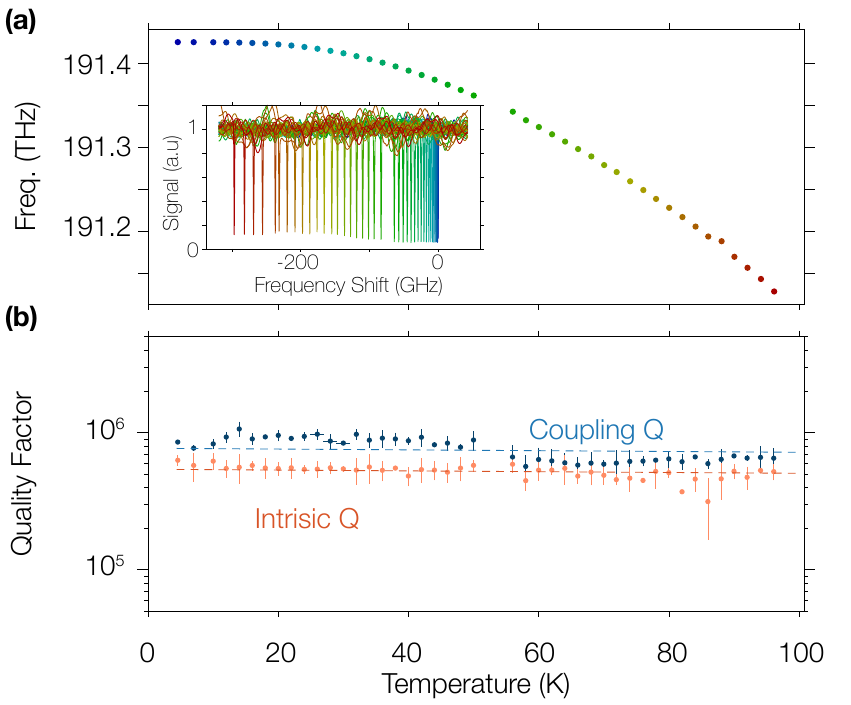}
    \caption{\label{fig:Linear} (a) Measured temperature-dependent shift in resonance frequency of the microring mode used to generate a soliton frequency comb. The inset displays the resonance with the trace color corresponding to the temperature. (b) Coupling quality factor (blue) and intrinsic quality factor (orange) of the pumped resonance. The uncertainty due to the fit is represented by the vertical bar and corresponds to one standard deviation.}
\end{figure}

Using the material dispersion relations and an eigeinfrequency solver, one can retrieve the dispersion profile and the integrated dispersion $D_\mathrm{int} = \nu_\mu - (\nu_{0} + \rm{FSR}\times$~$\mu )$ where $\mu$ represents the mode number relative to the pumped mode frequency $\nu_{0}$, and $\nu_\mu$ is the frequency of the $\mu$\textsuperscript{th} mode.The simulations in~\cref{fig:DispDint} show a region of anomalous dispersion suitable for supporting bright soliton states. In addition, $D_\mathrm{int}$=0 at $\approx$~270~THz, indicating that a dispersive wave (DW) can be phase-matched to the soliton at that frequency. However, the experimentally observed (and theoretically predicted) soliton bandwidth we observe is too narrow to enable measurable DW emission.

\subsection{Linear Measurements}

The linear measurements are performed using a wavemeter- and reference cell-based calibrated swept-wavelength laser system, allowing retrieval of the resonance frequency of the mode of interest (see \cref{fig:Linear}(a)) with $\approx$~0.1~pm accuracy. Using a doublet fitting model~\cite{borselli_beyond_2005}, we determine the intrinsic and coupling quality factors of the resonant modes.  We change the temperature of the sample holder in increments of 5~K, and the measured resonance frequency and extracted quality factors are shown in Fig.~\ref{fig:Linear}.

\subsection{Determination of thermo-refractive index}

\begin{figure}[h]
    \centering
    \includegraphics[width=\columnwidth]{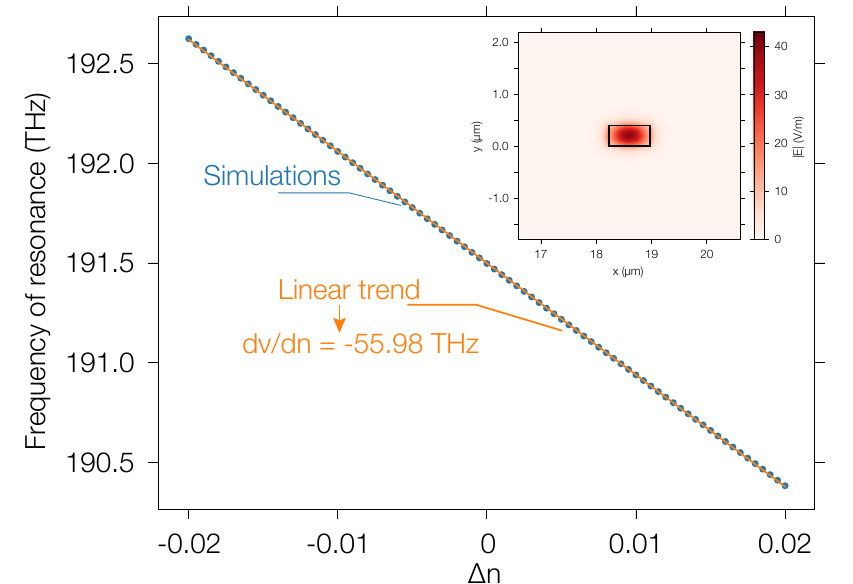}
    \caption{\label{fig:dfres_dn} Simulation of the shift of the microring resonance frequency with a small change of the refractive index of the microring core material $\Delta n$ (blue dots). The linear fit used to extract $\sfrac{\partial \nu}{\partial n}$ is represented through the orange line. The inset is a plot of the electric field amplitude of the mode of interest.}
\end{figure}

The thermo-refractive index of the system can be retrieved using the formula $\sfrac{\partial n}{\partial T} = \sfrac{\partial \nu}{\partial T} \left( \sfrac{\partial \nu}{\partial n} \right)^{-1}$, where  $\sfrac{\partial \nu}{\partial T} $ is obtained using the results presented in \cref{fig:Linear}(a), while $\sfrac{\partial \nu}{\partial n}$ is computed using an eigenfrequency solver. Here, we make the assumption that the influence of the SiO$_2$ cladding and underlying substrate is negligible.  This is justified for two reasons. First, according to the literature, the thermo-refractive coefficient of SiO\textsubscript{2} is low (between $5\times 10^{-7}$~K$^{-1}$ to $1 \times 10^{-6}$~K$^{-1}$) and its variation is negligible within the temperature range of interest. Second, the mode is highly confined within the AlGaAs core, as confirmed by comparing the mode area ($A_\mathrm{mode} = 1.94\times 10^{-13}$~m\textsuperscript{2}) to the ring resonator cross-section area ($A_\mathrm{ring} = 2.6\times 10^{-13}$~m\textsuperscript{2}. Hence, in the eigenfrequency simulation, we vary the refractive index of the AlGaAs and obtain $\frac{\partial \nu}{\partial n}$ from a linear fit (\cref{fig:dfres_dn}).

\subsection{Qualitative Determination of the Thermal Accessibility Region for Soliton States}

\begin{figure}[h]
    \centering
    \includegraphics[width=\columnwidth]{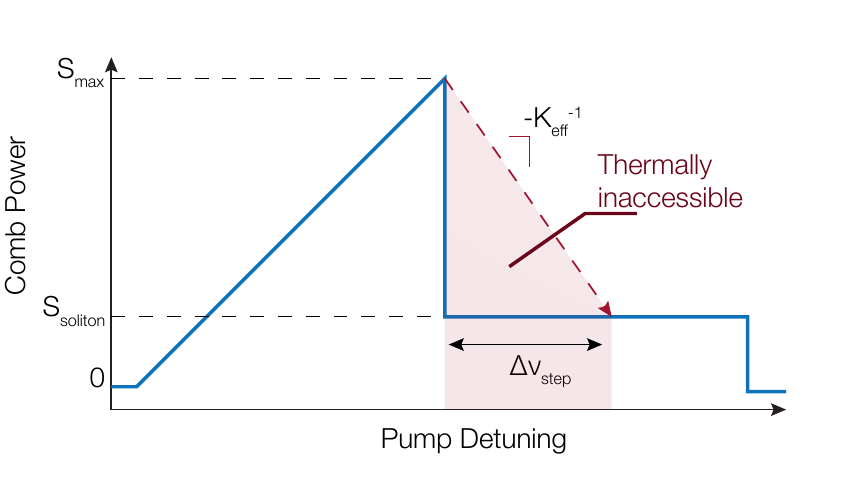}
    \caption{\label{fig:CartoonCombPower} Cartoon of the comb power (blue) taking into account only Kerr non-linearity (\textit{i.e.} not taking into account the thermal behavior). The red arrow correspond to the solution of the \textit{ad-hoc} equation presented in ref~\cite{Li2017}. Only the solution beyond the intersection of the Kerr-only solution and the linear thermal equation are thermally stable}
\end{figure}

From Li \textit{et al.}~\cite{Li2017}, the thermal accessibility window of the soliton states can be described by combining the LLE which describes Kerr dynamics by an auxiliary linear equation describing thermal dynamics. In~\cref{fig:CartoonCombPower}, we have represented the comb power (in blue) taking into account only the Kerr non-linearity (\textit{i.e.} pure LLE model) and displaying the signature soliton step. The soliton can only be thermally accessible beyond the intersection between the linear thermal equation (shown in dashed red) and the soliton step. Hence, the inaccessible pump-detuning region is:

\begin{equation*}
    \Delta \nu_\mathrm{step} = K_\mathrm{eff}^{-1}\left(S_\mathrm{max} - S_\mathrm{soliton} \right)
\end{equation*}

In first approximation, one could estimate the rise of the temperature through the resonator such that $H = mC_\mathrm{h}\Delta T$ with $H$ the heat (units of energy), $C_\mathrm{h} = 0.33+0.12x $~J$\cdot$g$\cdot$C\textsuperscript{-1} is the specific heat of Al\textsubscript{0.2}Ga\textsubscript{0.8} and $m$ its mass. Hence it becomes:

\begin{equation*}
    \Delta T = \frac{FSR \cdot Q_0 P_\mathrm{wg} \sfrac{\kappa_a}{\kappa} }{\rho V C_\mathrm{h}}
\end{equation*}

\noindent where $FSR = 617$~GHz is the free spectral range, $Q_0$ the intrinsic quality factor, $\kappa_\mathrm{a}$ and $\kappa$  are the absorption rate and the sum of the intrinsic and coupling rates respectively, $P_\mathrm{wg}$ is the in-waveguide power which we assume is entirely transferred to the resonator (\textit{i.e.} critical coupling regime), $V$ is the volume of the resonator, and $\rho=5.32-1.56x$~g$\cdot$cm\textsuperscript{-3} is the density of Al\textsubscript{0.2}Ga\textsubscript{0.8}As. From \cite{Li2017}, we further have:

\begin{equation*}
    K_\mathrm{eff}^{-1} = 2 \frac{\omega_0 t_\mathrm{r}}{n_\mathrm{g}K_\mathrm{c}}\frac{\kappa_\mathrm{a}}{\kappa} \frac{\partial n}{\partial T}\left(T_0 + \Delta T\right)
\end{equation*}

Using the above equations and assuming physically justified estimates for $S_\mathrm{max}$,  $S_\mathrm{soliton}$ and $\sfrac{\kappa_a}{\kappa}$, one is able to produce the qualitative thermal accessibility plots presented in Fig. 4 of the main manuscript.
\clearpage

\end{document}